\documentclass[10pt, a4paper]{article}
\usepackage{lrec}
\pdfoutput=1
\usepackage{multibib}
\usepackage{graphicx}
\usepackage{tabularx}
\usepackage{soul}
\usepackage{enumitem}

\usepackage{epstopdf}
\usepackage[utf8]{inputenc}
\usepackage{amssymb}
\usepackage{maltese}
\usepackage{hyperref}
\usepackage{xstring}

\usepackage{color}

\newcounter{example}
\newenvironment{example}[1][]{\refstepcounter{example}\par\medskip
   (\theexample) #1 \rmfamily}{\medskip}

\title{Annotating for Hate Speech: The MaNeCo Corpus and Some Input from Critical Discourse Analysis}

\name{Stavros Assimakopoulos, Rebecca Vella Muskat, Lonneke van der Plas, Albert Gatt}

\address{University of Malta \\
         Institute of Linguistics and Language Technology, Msida MSD 2080, Malta \\
         \{stavros.assimakopoulos, rebecca.vella, lonneke.vanderplas, albert.gatt\}@um.edu.mt\\}

\abstract{
This paper presents a novel scheme for the annotation of hate speech in corpora of Web 2.0 commentary. The proposed scheme is motivated by the critical analysis of posts made in reaction to news reports on the Mediterranean migration crisis and LGBTIQ+ matters in Malta, which was conducted under the auspices of the EU-funded C.O.N.T.A.C.T. project. Based on the realization that hate speech is not a clear-cut category to begin with, appears to belong to a continuum of discriminatory discourse and is often realized through the use of indirect linguistic means, it is argued that annotation schemes for its detection should refrain from directly including the label 'hate speech,' as different annotators might have different thresholds as to what constitutes hate speech and what not. In view of this, we suggest a multi-layer annotation scheme, which is pilot-tested against a binary $\pm$hate speech classification and appears to yield higher inter-annotator agreement. Motivating the postulation of our scheme, we then present the MaNeCo corpus on which it will eventually be used; a substantial corpus of on-line newspaper comments spanning 10 years.
\\ \newline \Keywords{hate speech, annotation, newspaper comments, corpus creation} }

\begin{document}

\maketitleabstract

\section{Introduction}

While the discussion of automatic hate speech detection goes back at least two decades \cite{Spertus97,Greevy2004}, recent years have witnessed a renewed interest in the area. This is largely due to the proliferation of user-generated content as part of Web 2.0, which has given rise to continuous streams of content, produced in large volumes over multiple geographical regions and in multiple languages and language varieties. As a result, the exponential increase in potential sources of hate speech in combination with the continuous introduction of legislation and policy-making aiming specifically at regulating the phenomenon across a number of countries \cite{Banksdoi:10.1080/13600869.2010.522323} clearly necessitates the development of reliable automatic methods for detecting hate speech online that will complement human moderation.

From a natural language processing (NLP) perspective, treatments of hate speech focus mainly on the problem of identification \cite{Fortuna:2018:SAD:3236632.3232676}. Thus, given a span of text, the task is to identify whether it is an instance of hate speech or not. This makes the problem a case of binary classification ($\pm$hate speech), which in turn makes it amenable to treatment using a variety of classification methods. These supervised learning techniques require pre-labelled training data, consisting of manually annotated positive and negative examples of the class(es) to be identified, to learn a model which, given a new instance, can predict the label with some degree of probability. In this setting, the most common features traditionally used for hate speech classifiers are lexical and grammatical features \cite{schmidt-wiegand-2017-survey,Fortuna:2018:SAD:3236632.3232676}, with more recent approaches making use of neural network models relying on word embeddings \cite{Badjatiya:2017:DLH:3041021.3054223,Agrawal10.1007/978-3-319-76941-7_11}.

In this paper, we concentrate on the particular issues that one has to take into consideration when annotating Web 2.0 data for hate speech. The unique angle of our perspective is that it is informed by data-driven research in the field of Critical Discourse Analysis (CDA), a strand of applied linguistics that has for long dealt with the ways in which language is used to express ideologically-charged attitudes, especially in relation to discrimination. For some of us the interest in the area of hate speech detection in the Web 2.0 era was originally sparked through our involvement in the C.O.N.T.A.C.T. project which specifically targeted online hate speech from a CDA point of view \cite{assimakopoulos_online_2017}. As a matter of fact, the ongoing compilation of the Maltese Newspaper Comments (MaNeCo) corpus, which this paper additionally launches, started off as a potential extension to the much smaller corpus that was compiled for the purposes of in-depth qualitative analysis undertaken under the auspices of C.O.N.T.A.C.T. \cite{assimakopoulos_exploring_2017,assimakopoulos_xenophobic_2018}. Against this backdrop, the ensuing discussion focuses on the challenges faced while annotating this corpus, which we are treating herein as a pilot for the development of the scheme used to annotate samples of MaNeCo, and which, given MaNeCo’s special characteristics, is expected to lead to more reliable datasets that could be used for future training and testing models of automatic hate speech detection. Thus, our main contributions are the following:
\begin{itemize}
\item A description of the challenges encountered when annotating hate speech 
\item An annotation scheme that aims to address these challenges
\item A pilot dataset of on-line newspaper comments on migration and LGBTIQ+ matters, with multi-level annotations (described in Section \ref{sec:pilot}).\footnote{The pilot dataset is available from the authors upon request.}
\item The MaNeCo corpus: a substantial corpus of Maltese newspaper comments spanning 10 years.\footnote{Samples of this corpus can be made available upon request. A full release is expected in future, pending licensing agreements with the donors of the MaNeCo data, which will need to cover sensitive data such as comments written by online users, but deleted by the moderators of the newspaper portal.}
\end{itemize}


\section{ Background}

When it comes to automatic hate speech detection, a sufficiency of training data with high-quality labelling is a crucial ingredient for the success of any model. Even so, previous studies “remain fairly vague when it comes to the annotation guidelines their annotators were given for their work” \cite[p.8]{schmidt-wiegand-2017-survey}. A review of the relevant literature reveals that the majority of previous attempts to annotate Web 2.0 data for hate speech involves simple binary classification into hate speech and non-hate speech \cite{Kwok:2013:LHD:2891460.2891697,Burnap2015,Djuric:2015:HSD:2740908.2742760,Nobata:2016:ALD:2872427.2883062}. There are of course notable exceptions where the annotation scheme involved was more or less hierarchical in nature. For example, in \newcite{warner-hirschberg-2012-detecting}, annotators were  tasked with classifying texts on the basis of whether they constitute hate speech or not, but were additionally asked to specify the target of said speech in the interest of distinguishing between seven different domains of hatred (e.g. sexism, xenophobia, homophobia, etc). Then, acknowledging that hate speech is a subtype of the more general category of offensive language and can thus often be conflated with it, \newcite{DavidsonICWSM1715665} asked annotators to label tweets in terms of three categories: hate speech, offensive but not hate speech, or neither offensive nor hate speech. Along similar lines, \newcite{zampieri-etal-2019-predicting} introduce an explicitly hierarchical annotation scheme that requested annotators to code tweets on three consecutive levels: (a) on whether they contain offensive language; (b) on whether the insult/threat is targeted to some individual or group; and (c) on whether the target is an individual, a group or another type of entity (e.g. an organization or event). Finally, an annotation framework which, like the one proposed here, takes into account the intricate nature of hate speech was developed by \newcite{sanguinetti-etal-2018-italian}. Here, annotators were asked to not only provide a binary classification of Italian tweets as hate speech or not, but also to grade their intensity on a scale from 0 to 4, and indicate whether each tweet contains ironical statements or stereotypical representations as well as how it fares in terms of aggressiveness and offensiveness.

Despite the apparently increasing interest in the area and the development of all the more sophisticated annotation methodologies, a major cause for concern when it comes to annotations used for model training and testing is that reliability scores are consistently found to be low pretty much across the board \cite{warner-hirschberg-2012-detecting,Nobata:2016:ALD:2872427.2883062,Ross2016,tulkens2016automated,waseem-2016-racist,bretschneider_detecting_2017-2,schmidt-wiegand-2017-survey,de-gibert-etal-2018-hate}. This effectively suggests that, even in the presence of more detailed guidelines \cite{Ross2016,malmasi:2018:profanity,sanguinetti-etal-2018-italian}, annotators often fail to develop an intersubjective understanding of what ultimately counts as hate speech, and thus end up classifying the same remarks differently, depending on their own background and personal views \cite{saleemetal,Salminen20188554954}.

The common denominator of the existing NLP literature on the matter seems to be that annotating for hate speech is in itself particularly challenging. The most pertinent reason for this seems to be the notorious elusiveness of the label ‘hate speech’ itself. Despite having been established in legal discourse to refer to speech that can be taken to incite to discriminatory hatred, hate speech is now often used as an umbrella label for all sorts of hateful/insulting/abusive content \cite{Brown2017}. This much is strikingly evident when one takes into account that most of the studies reviewed in NLP state-of-the-art reports on hate speech detection (including our discussion so far) formulate the question at hand using different – albeit interrelated – terms, which apart from hate speech variously include harmful speech, offensive and abusive language, verbal attacks, hostility or even cyber-bullying, among others. In this vein, as \newcite[p.78]{waseem-etal-2017-understanding} observe and exemplify, this “lack of consensus has resulted in contradictory annotation guidelines – some messages considered as hate speech by \newcite{waseem-hovy-2016-hateful} are only considered derogatory and offensive by \newcite{Nobata:2016:ALD:2872427.2883062} and \newcite{DavidsonICWSM1715665}.”

The apparent confusion as to how to adequately define hate speech is of course not exclusive to NLP research, but extends to social scientific \cite{Gagliardoneetal2015} and even legal treatments of the theme \cite{Bleich2014,Sellars2016}. Given this general confusion, it is certainly no surprise that annotators, especially ones who do not have domain-specific knowledge on the matter, will be prone to disagreeing as to how to classify some text in the relevant task, as opposed to other comparable annotation tasks. 

Discussing this issue within the more general context of detecting abusive language, \newcite{waseem-etal-2017-understanding} identify two sources for the resulting annotation confusion: (a) the existence of abusive language directed towards some generalised outgroup, as opposed to a specific individual; and (b) the implicitness with which an abusive attitude can often be communicated. Despite targeting abusive language in general, the resulting two-fold typology has been taken to apply to the more particular discussion of hate speech too \cite{elsherief-2018-318,MacAvaney2019,mulki-etal-2019-l,rizos_augment_2019}, rendering the lack of an explicit insult/threat towards an individual target in terms of their membership to a protected group more difficult to classify as hate speech than a remark that explicitly incites to discriminatory hatred. This much seems to be further corroborated by a recent study by \newcite{Salminen:2019:OHR:3295750.3298954} which revealed that, when  evaluating the hatefulness of online comments on a scale of 1 (not hateful at all) to 4 (very hateful), annotators agree more on the two extremes than in the middle ground. Quite justifiably, this suggests that it is easier to classify directly threatening or insulting messages as hate speech, rather than indirectly disparaging ones.

What transpires from a review of the literature then is that the difficulty in annotating for hate speech lies primarily in those instances of hate speech that appear to fall under the radar of some annotators due to the ways in which incitement to discriminatory hatred can be concealed in linguistic expression. For us, this is precisely the point where CDA can offer valuable insight towards developing schemes for hate speech annotation. That is because CDA specifically seeks, through fine-grained analysis and a close reading of the text under investigation, to uncover the “consciousness of belief and value which are encoded in the language – and which are below the threshold of notice for anyone who accepts the discourse as ‘natural’” (Fowler, 1991, p. 67). As we will now turn to show, an appreciation of hate speech as an ideology-based phenomenon can substantially inform NLP research in the area too.

\section{Pilot dataset}\label{sec:pilot}

For the purposes of the C.O.N.T.A.C.T. project we focused specifically on analysing homophobia and xenophobia in local (that is, Maltese) user comment discourse \cite{assimakopoulos_exploring_2017,assimakopoulos_xenophobic_2018}. In this respect, the Maltese C.O.N.T.A.C.T. corpus, which served as a pilot for the presently proposed annotation scheme, was built and annotated following the common methodology established across the C.O.N.T.A.C.T. consortium \cite[pp. 17-20]{assimakopoulos_online_2017}. The dataset was formed by scraping Maltese portals for comments found underneath news reports related to LGBTIQ+ and migrant minorities over two time periods: April-June 2015 and December 2015-February 2016. The identification of relevant articles was facilitated by the use of the EMM Newsbrief webcrawler\footnote{\url{http://emm.newsbrief.eu}}, where we performed a search for articles from Malta containing keywords pertaining to migrants and the LGBTIQ+ community in turn. We then scraped comments amounting to approximately 5,000 words worth of content per keyword, equally distributed across the selected keywords, eventually forming two subcorpora: one for migration, comprising 1130 comments (41020 words), and one for LGBTIQ+ matters, comprising 1109 comments (40924 words).

Following the compilation of the corpus, we engaged in annotation which comprised two steps. In the first instance, comments were classified in terms of their polarity, as positive or negative, depending on their underlying stance towards the minority groups in question. Then, through an in-depth reading of each comment labelled as negative, we identified the discursive strategies that underlie the communication of the negative attitude at hand. Crucially, this deeper level of annotation included the detection not only of linguistic forms, such as the use of derogatory vocabulary and slurs, tropes, or generics, but also implicit pragmatic functions that underlie the communication of insults, threats, jokes and stereotyping.

\section{The special nature of hate speech}

Right from the beginning of the aforementioned annotation of the pilot dataset, one issue that became immediately evident was that, much like in the majority of the corresponding NLP research, a shallow binary classification of positive and negative attitude cannot possibly do justice to the particularities of hate speech. This realisation appears to be in line with the argument made independently by the several NLP researchers who have proposed, as we have seen, more complex annotation systems than a simple $\pm$hate speech classification. Against this backdrop, the minute analysis performed on each comment for C.O.N.T.A.C.T. could not lend itself to multiple repetitions by different annotators in order to establish agreement, but it still reveals some principal reasons why the traditional reliance on lexical and grammatical features for the training of a hate speech detection model falls short of capturing the intricate nature of the category in question.

\subsection{Hate speech and offensive language}

As common experience indicates, the internet is full of emotional language that can be considered insulting and disrespectful towards both individuals and collective groups. In this setting, online hate speech might often contain offensive language, but not all offensive language used on the internet can be considered hate speech. That is because hate speech, in its specialised legal sense, is typically tied to the specific criterion of incitement to discriminatory hatred in the several jurisdictions where it is regulated, while the discriminatory attitude needs to specifically target a group that is legally protected.

Indeed, since our pilot corpus was annotated for hate speech specifically targeting migrants and the LQBTIQ+ community, it became obvious that there were a number of comments aimed at other groups of people as well as individuals too. As the following response from one commenter to another makes clear, a simple positive-negative attitude classification could easily lead to the inaccurate labelling of data within the corpus:

\begin{example}
\textit{I'm not your Hun pervert!}
\end{example}

While this comment is not targeted at a minority, it is a very direct insult, unlike much of the discourse that pointed to minorities in our dataset, which was more often than not only indirectly offensive. Similarly, although less abusive, the following comment also offends another online user: 

\begin{example}
\textit{I hope you're just trolling [username]. If not, you truly are a sad being.}
\end{example}

Clearly, despite being obviously offensive, the two examples do not constitute hate speech, since they are not targeted toward any group, much less a protected minority.  

In a similar vein, both (3) and (4) do not take issue with a minority group, but rather target groups that can be seen as dominant in the Maltese context.

\begin{example}
\textit{The issue or problem is that religion and government are not truly separate on Malta and never were or will be in the future no matter what is written in any constitution.}
\end{example}

\begin{example}
\textit{Just remember that they wouldn't be working Illegaly if there wasn't someone willing to employ them Illegaly, and a lot of these employers are Maltese.}
\end{example}

Evidently, (3) criticises the fact that the church and state in Malta are still intertwined, while (4) undermines the general negative stance taken toward migrants by expounding the hypocrisy of such a stance alongside the willingness of the Maltese to hire members of this group and thus benefit from cheap labour. Again, within the framework of discriminatory discourse, although these examples display a negative disposition against collective groups (church, state, the Maltese), they cannot be deemed hate speech, since the groups targeted are not protected under Maltese Law.

\subsection{Hate speech as a continuum}
Beyond the examples which show that not all comments labelled as negative constitute hate speech in the legislative sense of the word, a mere label of ‘negative’ fails to capture the complexity of the scale on which various forms of incitement to discriminatory hatred fall. In this respect, the analysis of the C.O.N.T.A.C.T. corpus also revealed that discriminatory hatred can vary from direct calls to violence, as in (5), to indirect forms of discrimination, like the one exemplified in (6):

\begin{example}
\textit{They could use a woman to execute them because they believe that if they are killed by a woman they will not go to heaven and enjoy their 77 virgins.}
\end{example}

\begin{example}
\textit{I believe the problem is Europe wide, but I feel Malta is harder hit simply because of it's size. Malta simply has not got the room to keep receiving these people.}
\end{example}

While the commenter in (5) suggests that Muslim migrants are executed, the one in (6) does not appear at first sight to articulate anything directly offensive or insulting; upon closer inspection, however, it can be taken to allude to ideas of exclusion and in-grouping, since the use of ‘these people’ serves to create an in-group of \textit{us} (who belong here and deserve access to resources) and \textit{them} (who do not belong here and should not be taking up our precious space). 

While the two examples given above illustrate the two opposite ends of a spectrum, there is of course much that lies in between. In the following comment, for example, the commenter may acknowledge the need to respect members of the LGBTIQ+ community, but concurrently refers to specific members of the minority group, that is transgender individuals, as ‘too complicated and abnormal:’

\begin{example}
\textit{We just need to teach our children to respect each other whoever they be. Teaching gender diversity is too complicated and abnormal for their standards. I’m afraid it would effect their perception that lgbti is the norm instead of a minority.}
\end{example}

So, despite emphasising the importance of promoting diversity, the commenter also recommends that children are taught to differentiate between the ‘normal’ dominant identities and ‘abnormal’, and thus subordinate ones. Equivalently, the commenter in [8] explicitly denies racism, while at the same time clearly employing tactics of negative stereotyping and categorisation with the use of the term ‘illegals.’

\begin{example}
\textit{[username] sure NO! The majority of Maltese people are against " ILLEGALS"! Do not mix racism with illegals please!}
\end{example}

All in all, as the examples above show, there are varying degrees of discriminatory hatred that need to be accommodated within the purview of hate speech. Crucially, this is something that is underlined in non-NLP research on hate speech too (Vollhard, 2007; Assimakopoulos et al., 2017). Cortese (2007, p. 8-9), for example, describes a framework that treats hate speech not as a single category, but rather as falling on a four-point scale:
\begin{enumerate}
\item \textit{unintentional discrimination}: unknowingly and unintentionally offending a (member of a) minority group by, for example, referring to the group by means of a word that is considered offensive or outdated, such as referring to black people as ‘coloured’ or by referring to asylum seekers as ‘immigrants;’
\item \textit{conscious discrimination}: intentionally and consciously insulting a (member of a) minority group by, for example, using a pejorative term like ‘faggot’ or by describing to migrants as ‘invaders;’
\item \textit{inciting discriminatory hatred}: intentionally and consciously generating feelings of hatred toward minorities by publicly encouraging society to hate and exclude the group in question, such as by suggesting that members of the LGBTIQ+ community are sick or that migrants bring contagious diseases from their countries;
\item \textit{inciting discriminatory violence}: intentionally and consciously encouraging violence against minorities by, for example, suggesting that members of a minority group be executed.
\end{enumerate}

The importance of a micro-classification of our negative comment data becomes apparent against the backdrop of Cortese’s categorisation, since not all the examples given above would fall within the third or fourth regions of the scale, which are the points that most current hate speech legislation appears to regulate. That said, given the wider use of the term hate speech in everyday discourse, several annotators could take texts that fall under the first two points of Cortese’s scale to constitute hate speech too.

\subsection{Hate speech and implicitness}
The discussion so far inevitably leads to what we consider to be the main challenge in achieving an intersubjective understanding of hate speech. Clearly, explicit incitement, of the type expressed by an utterance of “\textit{kill all [members of a minority group]}” or “\textit{let’s make sure that all [members of a minority group] do not feel welcome},” is not very difficult to discern. However, since “most contemporary societies do disapprove of” such invocations, “overt prejudicial bias has been transformed into subtle and increasingly covert expressions” \cite[p.146]{Leetsdoi:10.1177/0261927X03022002001}. Therefore, while clearly indicative of a negative attitude, the use of derogatory terms, as in (9), cannot account for all instances of hate speech:
\begin{example}
\textit{That’s because we're not simply importing destitute people. We're importing a discredited, disheveled and destructive culture.}
\end{example}

In this respect, it is often acknowledged \cite{warner-hirschberg-2012-detecting,gao-huang-2017-detecting,schmidt-wiegand-2017-survey,Fortuna:2018:SAD:3236632.3232676,sanguinetti-etal-2018-italian,Watanabe8292838} that discourse context plays a crucial role in evaluating remarks, as there are several indirect strategies for expressing discriminatory hatred, which our in-depth analysis enabled us to additionally unearth. The most pertinent example of this is the use of metaphor, which has for long been emphasised as a popular strategy for communicating discrimination, since it typically involves “mappings from a conceptual ‘source domain’ to a ‘target domain’ with resulting conceptual ‘blends’ that help to shape popular world-views in terms of how experiences are categorized and understood” \cite[p.24]{Musolffdoi:10.1080/00313220601118744}. In (10), for example, the commenter engages in incitement by making use of the metaphorical schema \cite{Lakoff2003} {\sc migration is a disease}: 
\begin{example}
\textit{Illegal immigration is a cancer which if not eliminated will bring the downfall of Europe and European culture.}
\end{example}

Similar indirect strategies can be found in the frequent use of figurative language to underline the urgency of the situation, as exemplified by the use of allusion in (11) and hyperbole in (12):
\begin{example}
\textit{Will Malta eventually become the New Caliphate?}
\end{example}
\begin{example}
\textit{\textellipsis in 4 more days we will become the minority \textellipsis}
\end{example}

Alongside figurative language, stereotypical representations and remarks generalising over a minority group, as in (12) also play a crucial role in the expression of discriminatory hatred \cite{Brown,haas2012,Maitra,Kopytowska2017}:
\begin{example}
\textit{If anyone is lacking, it is you guys for lacking a sense of decency. Just look at the costumes worn at gay parades to prove my point.}
\end{example}

Now, while such remarks might not appear to fall under the incitement criterion for the delineation of hate speech, there is still reason to include them in a corpus of hate speech. As \newcite[p.85:5]{Fortuna:2018:SAD:3236632.3232676} argue, “all subtle forms of discrimination, even jokes, must be marked as hate speech” because even seemingly harmless jokes indicate “relations between the groups of the jokers and the groups targeted by the jokes, racial relations, and stereotypes” \cite{Kuipers2016} and their repetition “can become a way of reinforcing racist attitudes” \cite{Kompatsiarisdoi:10.1080/13504630.2016.1207513} and have "negative psychological effects for some people" \cite{Douglass2016}. 

Be that as it may, people are bound to disagree as to whether such implicit expressions can or should indeed be classified as hate speech. In view of this, we think that a strategy for obtaining more reliable annotation results would be to refrain as much as possible from asking raters to make a judgement that could be influenced by their subjective opinion as to what ultimately constitutes hate speech, as well as by their own views and attitudes towards the minority groups under question. 

\section{Towards a new annotation scheme}\label{section-annotationscheme}

We trust that our discussion so far provides insight as to why it is particularly difficult to achieve adequate agreement among different crowd coders when it comes to classifying hate speech. 
Given that non-experts cannot always distinguish between hate speech and offensive language, and given the varying thresholds that hate speech laws place on the continuum of discriminatory discourse,
it would probably be hard even for hate speech experts to fully agree on a classification \cite{waseem-2016-racist,Ross2016}. It is thus clear that a simple binary classification of online posts into hate speech or non-hate speech is unlikely to be reliable, and that a more complex scheme is inevitably needed.
When we presented previous annotation frameworks above, we mentioned that the one developed by \newcite{sanguinetti-etal-2018-italian} was informed by critical discussions of the concept of hate speech and was thus relatively complex. While on the right track, however, it seems to also incorporate categories, such as intensity, aggressiveness and offensiveness, that are susceptible to a rather subjective evaluation. Indispensable though they may be for the discussion of hate speech, such categories can easily compromise agreement as well, since individuals with different backgrounds and views could provide markedly different interpretations of the same data.

The proposal that we wish to make in this paper is that such subjective categories are left out from annotation instructions as much as possible. To this effect, the scheme that we developed for hate speech annotation in the MaNeCo corpus is hierarchical in nature and expected to lead to the identification of discriminatory comments along the aforementioned scale by Cortese without focusing on such impressionistic categories as intensity or degree of hatefulness, as follows:

\begin{enumerate}
\item Does the post communicate a positive, negative or neutral attitude? [\textbf{Positive / Negative / Neutral}]
\item If negative, who does this attitude target? [\textbf{Individual / Group}]
\begin{enumerate}
\item  If it targets an individual, does it do so because of the individual’s affiliation to a group? [\textbf{Yes / No}] If yes, \textbf{name the group}.
\item  If it targets a group, \textbf{name the group}.
\end{enumerate}

\item  How is the attitude expressed in relation to the target group? Select all that apply. [\textbf{ Derogatory term / Generalisation / Insult / Sarcasm (including jokes and trolling) / Stereotyping / Suggestion / Threat }]\footnote{The selection of these particular communicative strategies as opposed to alternative ones is based on the categories used for the purposes of the C.O.N.T.A.C.T. project. We thus assume an adequate coverage, since all these strategies were identified in the parallel annotation of corpora from the nine different national contexts represented in the project. That being said, we considered adding a field for annotators to suggest their own categories, but decided against it, since this could easily lead to confusion when it comes to grouping different categories together during the processing of responses.}
\begin{enumerate}
\item If the post involves a suggestion, is it a suggestion that calls for violence against the target group? [\textbf{Yes / No}]
\end{enumerate}
\end{enumerate}

With regards to the proposed annotation scheme, we cannot of course fail to acknowledge that the first label (positive, negative, neutral) is to a certain extent subjective. Still, we believe that it is formulated in a way that does not require the annotator to make a judgement in relation to a post’s hatefulness or aggressiveness, but merely to evaluate the commenter’s attitude in writing the comment. Obviously, this would inevitably lead to some disagreement, but it ultimately calls for a judgement that should be more straightforward to make, as it is not tied to an individual's understanding of what constitutes discrimination {\em per se}; after all, a negative attitude can easily crop up in various other settings too, like, for example, when expressing disagreement to someone else’s post in the same thread.

So, once negative attitude is established, the second step in the process would help to indirectly assess whether the attitude can be taken to be discriminatory too, insofar as it targets a group or an individual on the basis of group membership. In this vein, depending on whether the group that each discriminatory post targets is protected by law – like migrant or LGBTIQ+ individuals usually are, but politicians or church officials are not – we can distinguish between hate speech and merely hateful discourse. 

Finally, the third and last step will help determine the positioning of a post that fulfils the penultimate and final criteria in Cortese’s scale. In this regard, posts that comprise a suggestion would fall under incitement to discriminatory hatred, while posts specifically calling for violent actions towards the target group and those including threats would belong to the category of incitement to discriminatory violence. By the same token, posts containing insults, derogatory terms, stereotyping, generalisations and sarcasm would fall under Cortese’s categories of conscious and unintentional discrimination, which are not – at face value – regulated by hate speech law, but are still, as we have seen, relevant to the task at hand. In this way, one should be able to establish different sets of discriminatory talk, which could then be included or excluded from subsequent analyses, depending on the threshold that one specifies for the delineation of hate speech. 

Apart from enabling us to distinguish between explicit hate speech and softer forms of discrimination, a further merit of the proposed annotation scheme is that it indirectly allows us to control for annotator disagreement in some cases. The most obvious such case would be the presence of posts that are ambiguous between a literal and a sarcastic interpretation, or even the identification of lexical items and generalisations that might be considered offensive by some annotators and not by others. At the same time, it could provide useful indications regarding the distinction between conscious and unintentional discrimination. The rule of the thumb here would be that the more annotators agree on a comment belonging to the first two categories of discrimination in Cortese's scale, in the sense that it is discriminatory but does not include a suggestion regarding the target group, the closer that comment would be to conscious discrimination. That is because, as we have seen, unintentional discrimination has a marked tendency to go unnoticed and is thus expected to be less noticed by the annotators. Crude though this generalisation might be, we believe that it still provides a criterion that is viable.\footnote{An alternative here would potentially involve some kind of user profiling that would allow for an identification of the commenter's more general stance towards the target minority. Such an alternative, however, could compromise the overall task, since grouping a user's comments together is bound to bias annotators.} 

Obviously, this is not to say that disagreement can be completely eradicated. As a matter of fact, we cannot emphasise enough that the aim of our suggested annotation scheme (or of any other such scheme for that matter) is not to achieve perfect agreement. After all, as our preliminary CDA analysis revealed, genuine ambiguity can present itself not only at the structural, but also at the attitudinal level. For example, the following comment was posted in reaction to an article about a Nigerian rape victim who was fighting to bring one of her two children to Malta: 
\begin{example}
\textit{For God’s sake, make the process simpler and allow this woman to unite with her daughter in Nigeria.}
\end{example}

At face value, it seems difficult to discern if the commenter wishes for the woman in question, who is stranded in Malta due to not having received refugee statues, to be allowed to reunite with her daughter by bringing her to Malta, or is suggesting that she be deported back to Nigeria in order to be with her daughter. Apart from this, some posts may still be difficult to classify on the ground that they express multiple attitudes. For example, in stating (15), the commenter acknowledges that marriage should be a matter of choice between two people in love, but then directly implies that marriage should be a union between two people of the opposite (cis)gender. 
\begin{example}
\textit{People marry because they fall in love, and although it’s a choice, it was meant to be like that even in the animal kingdom, for example swans mate for life, male and female, not male and male.}
\end{example} 

Despite the inevitable presence of such borderline ambiguous comments, however, we do expect our annotation scheme to fare better in terms of inter-annotator agreement than previous attempts, and particularly attempts based on a binary $\pm$hate speech classification. In an attempt to assess its efficacy then, we conducted a preliminary pilot study to which we will now turn.

\subsection{Piloting the new scheme}

A total of 24 annotators took part in this pilot study. The participants, who were mostly academics and students ranging between 21 and 60 years of age, were divided into two gender- and age-balanced groups of 12. The first group was asked to label items using simple binary annotation ($\pm$hate speech) on the basis of the definition of hate speech provided within the Maltese Criminal Code, while the second used our proposed multi-level annotation scheme. Both groups were presented with 15 user-generated comments from the Maltese C.O.N.T.A.C.T. corpus in random order. In an effort to ensure variation of the items to be annotated, we went back to our original CDA classification and selected three comments from each of the following categories:
\begin{itemize}
\item comments involving incitement to discriminatory violence against the migrant minority;
\item comments that were labeled as discriminatory towards the migrant minority but not as fulfilling the incitement criterion;
\item comments that were labeled as negative but do not target the migrant minority; 
\item comments that were labeled as expressing a positive attitude towards the migrant minority; and
\item comments that were labeled as ambiguous, along similar lines to the discussion of (14) and (15) above.
\end{itemize}
All individuals in both annotation conditions performed the task independently. Obviously, in order to meaningfully compare the levels of inter-annotator agreement for the two schemes, we needed to have an equal number of classes across the board. To achieve this we screened the annotations received for the multi-level condition, with a view to inferring a binary $\pm$hate speech classification in this case too. 

While inferring a binary class from the multi-level scheme might seem counter-intuitive, in view of the preceding discussion, there are several reasons why we did this. First, the decision was taken on pragmatic grounds: in order to achieve a fair comparison, agreement within the two groups needed to be estimated based on similar categories. That way, we were able to compute inter-annotator agreement between participants in the two annotation tasks in a way that would enable a direct comparison. More importantly, however, there are theoretical grounds for using the multi-level scheme to achieve a binary classification. One of the arguments in the previous section was that hate speech would be evident in comments that would be labeled as negative, targeting (members of) a minority group and comprising either a suggestion or a threat. Indeed, the thrust of the argument presented above is not that a binary classification is undesirable, but that in order to be reliably made, it had to supervene on micro-decisions that took into account these various dimensions. The extent to which this is the case is of course an empirical question, one to which we return in the concluding section.

In Table \ref{table-agreement}, we report percent agreement, Fleiss's kappa \cite{Fleiss-71}, and Randolph's kappa \cite{Randolph05}, which is known to be more suitable when raters do not have prior knowledge of the expected distribution of the annotation categories. All in all, the results from this pilot appear to corroborate our prediction: the proposed multi-level annotation scheme appears to indeed lead to higher inter-annotator agreement, which for Fleiss's kappa in particular seems to even mark a rise from moderate (0.54) to substantial agreement (0.69).

\begin{table}\label{table-agreement}
\begin{center}
\begin{tabular}{lrr}
\hline 
Metric          & binary    & multi-level\\
\hline
Percent agr.    &76.8\%     &84.6\% \\
Fleiss' $k$     &0.54       &0.69 \\
Randolph's $k$  &0.48       &0.58 \\
\hline
\end{tabular}
\end{center}
\caption{Inter-annotator agreement for the two annotation schemes}
\end{table}

Apart from enabling us to indirectly compare its performance in relation to the corresponding binary classification scheme, the proposed multi-level scheme also allowed us to pinpoint those levels at which disagreement was more pronounced. So, whereas participants agreed more when asked about the positive/negative/neutral attitude of the speaker, as well as about the  individual/group targeted by a comment, they agreed far less when asked to evaluate more concretely how a negative attitude is expressed in each comment. Informal discussions with the participants upon completion of the task suggested that this was due to their lack of experience with discerning the relevant discourse strategies, but additionally shed light on one further source of disagreement: some annotators expressed difficulty in distancing themselves from the attitude of the commenter, and felt tempted to rate a comment as negative if the commenter expressed an attitude opposing their own. Clearly, this is valuable feedback about where we would need to focus our attention while further improving the annotation scheme and accompanying instruction text.

\section{Future steps: The MaNeCo corpus}

Having motivated the proposed annotation framework on the basis of our pilot study, we are currently in the process of implementing it in the annotation of various samples for the MaNeCo corpus. At the moment, the MaNeCo corpus comprises original data donated to us by the \textit{Times of Malta}, the newspaper with the highest circulation in Malta. More specifically, it contains -- in anonymised form -- all the comments from the inception of the Times of Malta online platform (April 2008) up to January 2017 (when the data was obtained). This amounts to over 2.5 million user comments (over 124 million words), which are written in English, Maltese or a mixture of the two languages, even though the newspaper itself is English-language. Our aim is to eventually populate it with data from other local news outlets too to ensure an equal representation of opinions irrespective of a single news portal’s political and ideological affiliations. That said, even this dataset on its own is an invaluable resource for hate speech annotation, since it also includes around 380K comments that have been deleted by the newspaper moderators, and which are obviously the ones that tend to be particularly vitriolic. In this regard, seeing that there are generally “much fewer hateful than benign comments present in randomly sampled data, and therefore a large number of comments have to be annotated to find a considerable number of hate speech instances” \cite[p.7]{schmidt-wiegand-2017-survey}, MaNeCo is well suited for addressing the challenge of building a set that is balanced between hate speech and non-hate speech data. 

Clearly, this does not mean that we will not have additional challenges to face before we end up with a dataset that could potentially be used for training and testing purposes. For one, being a corpus of user-generated content in the bilingual setting of Malta, it contains extensive code-switching and code-mixing \cite{Rosner2007ATA,elfardy-diab-2012-token,sadat-etal-2014-automatic,eskander-etal-2014-foreign}, which is further complicated by the inconsistent use of Maltese spelling online, particularly in relation to the specific Maltese graphemes 
[{\em ċ}], [{\em ġ}], [{\em g\mh}], [{\em \mh}], and [{\em ż}],
for which diacritics are often omitted in online discourse. Then, given the casual nature of most communication on social media, it is full of non-canonical written text \cite{baldwin-etal-2013-noisy,eisenstein-2013-bad} exhibiting unconventional orthography, use of arbitrary abbreviations and so on. 

Even so, these are challenges that can be faced after the effectiveness of our proposed scheme in yielding better inter-annotator agreement results is established more concretely. In this regard, a line of future work, briefly discussed in Section~\ref{sec:pilot}, is to further validate the multi-level annotation schemes. Specifically, reliability needs to be estimated on the basis of larger and more diverse samples. Furthermore, we have already identified the question of whether, having conducted a micro-analysis of user-generated texts, it becomes easier to classify them, in a second step, as instances of hate or non-hate speech, thereby going from a multi-level to a binary classification. The results from our pilot study are encouraging, in that they evince higher agreement when annotators have their attention drawn to different aspects of the text in question. Whether this will also translate into a more reliable detection rate for hate speech -- by humans or by classification algorithms, for which the multi-level annotation may provide additional features -- is an empirical question we still need to test.

In closing, to the extent that we can assess our proposed annotation scheme's usefulness for training automatic classifiers, we believe that, by distinguishing between different target groups from the beginning, this scheme could prospectively enable us to select training material from different domains of hate speech (such as racism, sexism, homophobia, etc.) in a way that serves transfer learning too. This could be particularly useful in addressing the notorious challenge of retaining a model’s good performance on a particular dataset source and/or domain of hate speech during transfer learning to other datasets and domains \cite{Agrawal10.1007/978-3-319-76941-7_11,Arango:2019:HSD:3331184.3331262}. After all, as \newcite{Grondahl:2018:YNL:3270101.3270103} concluded, after demonstrating – through replication and cross-application of five model architectures – that any model can obtain comparable results across sources/domains insofar as it is trained on annotated data from within the same source/domain, future work in automatic hate speech detection “should focus on the datasets instead of the models,” and more specifically on a comparison of “the linguistic features indicative of different kinds of hate speech (racism, sexism, personal attacks etc.), and the differences between hateful and merely offensive speech.” \cite[p.11]{Grondahl:2018:YNL:3270101.3270103}. This is a direction that we are particularly interested in, since our CDA research on the Maltese C.O.N.T.A.C.T dataset correspondingly revealed that although users discussing the LGBTIQ+ community and migrants appear to employ similar tactics in expressing discriminatory attitudes, the content of their utterances differs to a considerable extent depending on the target of their comments. \footnote{For example, while migrants are often referred to as ‘ungrateful’ or ‘dangerous’ and migration is typically framed in terms of metaphors of invasion, members of the LGBTIQ+ community are characterised as ‘abnormal’ and ‘sinners’ with frequent appeal to metaphors of biblical doom. }

\section{Acknowledgements}

The research reported in this paper has been substantially informed by the original work conducted by the University of Malta team on the C.O.N.T.A.C.T. Project, which was co-funded by the Rights, Equality and Citizenship Programme of the European Commission Directorate-General for Justice and Consumers (JUST/2014/RRAC/AG). We gratefully acknowledge the support of the Times of Malta in making the MaNeCo data available to us. Last but not least, we would like to thank Slavomir Ceplo for his diligent work in extracting and organising the MaNeCo corpus, the three anonymous LREC reviewers for their insightful comments and, of course, all participants of the pilot study for their generous availability.

\section{Bibliographical References}\label{reference}

\bibliographystyle{lrec}
\bibliography{lrec2020Assimetal}

\end{document}